\documentclass[prd,aps,floats,twocolumn,nofootinbib]{revtex4-1}
\usepackage{slashed}
\usepackage{mathtools}
\usepackage{amsfonts}
\usepackage{amssymb}
\usepackage{epsfig}
\usepackage{bm}
\usepackage{makecell}
\usepackage[dvipsnames]{xcolor}

\begin{document}

\newcommand{\m}[1]{\mathcal{#1}}
\newcommand{\nn}{\nonumber}
\newcommand{\ph}{\phantom}
\newcommand{\eps}{\epsilon}
\newcommand{\be}{\begin{equation}}
\newcommand{\ee}{\end{equation}}
\newcommand{\bea}{\begin{eqnarray}}
\newcommand{\eea}{\end{eqnarray}}
\newtheorem{conj}{Conjecture}

\newcommand{\plk}{\mathfrak{h}}


\title{Thermodynamics in space-times without horizons}
\date{}

\author{Raymond Isichei}
\email{raymond.isichei21@imperial.ac.uk}
\author{Jo\~{a}o Magueijo}
\email{magueijo@ic.ac.uk }
\affiliation{Abdus Salam Centre for Theoretical Physics, Imperial College London, Prince Consort Rd., London, SW7 2BZ, United Kingdom}

\begin{abstract}
We consider the energetics and thermodynamics of spacetimes with no horizons, but endowed with a preferred timelike junction surface. They could arise as a limiting case of the gravastar and other constructions regularizing the interior of the horizon of a black hole,  or  from the conceptual cutting of a portion of a non-asymptotically flat space and gluing it with flat space. We find that such surfaces can be made to have zero energy, so that the energetics of such spaces is not encumbered by them. They do have a transverse pressure, fixed by the jump in surface gravity. A peculiar matter thermodynamics then follows, with well defined entropy,  temperature and surface pressure, constrained by specific relations arising from the zero energy condition. This is confirmed by the Euclidean path integral, with the proviso that the Tolman-Ehrenfest temperature should be used. The entropy of the space is then the area of the time-like surface in units of a fundamental area, and the matter temperature is proportional to the transverse pressure and so the jump in surface gravity. However, when the gravitational action is added, the free energy of the surface is also zero.  The fact that for some such spaces the temperature comes out negative raises interesting questions regarding the third ``law'' of thermodynamics.


\end{abstract}

\maketitle

\section{Introduction}

Ever since Bekenstein proposed that black holes have entropy \cite{JB1, JB2, JB3}, the fact that some spacetimes have well defined thermal properties has been seen as a major clue informing the unification of gravity and quantum mechanics. There are, however, limits to such a ``thermal link''. The usual definitions of temperature and entropy in curved space-time assume the existence of a horizon, and are stated in terms of geometrical properties of this null surface \cite{H1,H2}. This leaves out important situations, in particular those arising from the drive to remove or smooth out singularities, with the horizon typically also excised as collateral (as in~\cite{gravastar1,gravastar2}, for example) or even as an essential part of the process\footnote{The view has been stated that removing the singularity without removing the horizon is like treating the symptoms of a disease rather than the disease. We thank Prof. Afshordi for this perceptive remark.}.

In many such cases, one finds that although there is no horizon, the spacetime is nonetheless endowed with a  preferred non-null surface, even if only conceptually. For example, if a static spacetime is not asymptotically flat, one may cut out a finite portion and glue it with empty space; or if we want to remove the singularity of a black hole we may excise the whole of the interior of its horizon and refill it with a portion of deSitter space. In all of these cases there is a junction, providing a preferred surface which is imelike rather than null \cite{gravastar1,gravastar2}.
In this paper we propose a definition for temperature, entropy and other thermodynamic quantities in space-times without horizons, but which are provided (or can be made to be) with a special 2+1 timelike surface.

Specifically, we restrict ourselves to static, spherically symmetric situations, typically not asymptotically flat. We then consider whether we can define mass, temperature and entropy for the space-time inside a given radius $r$ by cutting such a region and pasting it onto an empty outside space, with a joining surface which is typically timelike, not null. This can be seen as the basic motivation; however we will deal with the problem in complete generality, assuming just the gluing on an inside and an outside, and the energetic and thermodynamic properties of the timelike junction. 

In Section~\ref{Sec:energy}, 
we start by considering junction conditions between two static spherically symmetric geometries. We draw inspiration from the gravastar construction \cite{gravastar1,gravastar2}, in a variation where instead of allowing for a transition region between a regular non-empty interior and an empty exterior, we postulate an infinitely thin connecting surface. We find the remarkable result that in general such a surface has no energy. Hence it does not contribute to the Misner-Sharp energy of the system, and we can define the energy of the interior (and of the whole space-time) from the asymptotic ADM mass, unencumbered by the joining surface. 

This settles the issue of how to define the energy of such space-times, but what about their thermodynamics? Usually energy and temperature are closely related, but here the surface has no energy. In Section~\ref{thermodynamics} we examine the unorthodox thermodynamics of 2+1 surfaces which have no energy, but have pressure. The upshot is a Euler-like relation constraining the product of the matter entropy density and temperature to equal the surface pressure derived from the junction conditions. The transverse pressure is also forced to be proportional to the temperature, as a result of the Maxwell relations for the system. 

Obviously this leaves out gravity, so we re-examine the situation from the point of view of the Euclidean path integral in Section~\ref{Sec:Z}. We find that consistency requires that the definition of temperature to be adopted be the Tolman-Ehrenfest temperature. This was first proposed in~\cite{TE} and has been revived, for example, by Brown and York \cite{BY} and by \cite{CR} via the concept of ``temperature as the speed of time''. We confirm the matter thermodynamical results in Section~\ref{thermodynamics}, specifically that the entropy of the surface is its area and is independent of the temperature, and that the (Tolman-Ehrenfest) temperature is proportional to its transverse pressure (and so to the jump in surface gravity). However, when we add gravity to the path integral we obtain a vanishing free energy. We conclude with a discussion of these results. 

Throughout this paper we shall use units where $G=\hbar=c=k_B=1$.

\section{Energy in spaces with a timelike joint}\label{Sec:energy}
We base our analysis on Israel's junction conditions \cite{Israel} (or e.g.~\cite{Poisson}). We consider an internal geometry $\cal{M}_{(-)}$ with metric $g_{\mu \nu (-)}$ which induces  metric $\gamma_{ij (-)}$ on a constant $r=r_{0}$ time-like hypersurface $\Sigma$,  acting as a junction surface to 
corresponding external structures $\cal{M}_{(+)}$, $g_{\mu \nu (+)}$ and $\gamma_{ij (+)}$. 
The difference of tensorial quantities across the common hypersurface $\Sigma$ is of central importance when considering the joining of geometries. The crucial quantities are
$\big[\gamma_{ij}\big]\equiv \gamma_{ij(+)}-\gamma_{ij(-)}$, 
$\big[n_{\mu}\big]\equiv n_{\mu(+)}-n_{\mu(-)}$ and
$\big[k_{ij}\big]\equiv k_{ij(+)}-k_{ij(-)}$, 
where $k_{ij}$ is the extrinsic curvature tensor
and $n_{\mu}$ is the normal vector of $\Sigma$. 
Besides the continuity of the normal vector across the hypersurface ($\big[n_{\mu}\big]=0$) the relevant equations associated with the hypersurface are the Israel junction conditions
\begin{align}
        &\text{(J1)} \quad \big[\gamma_{ij}\big]=0,\label{J1}\\
        &\text{(J2)} \quad S_{ij}=-\frac{\epsilon}{8\pi}\bigg(\big[k_{ij}\big]-\big[k\big]\gamma_{ij}\bigg),\label{J2}
\end{align}
where $n_{\mu}n^{\mu}=\epsilon$ (with $\epsilon=1$ for the $£\Sigma$ contemplated in this paper). The condition (J1) enforces the commonality of the induced metric $\gamma_{ij}$ on the hypersurface $\Sigma$ when approached from either direction and (J2) attributes discontinuities in its normal derivative across the surface to a stress energy tensor $S_{ij}$ whose support lies solely on $\Sigma$. In this paper we take:
\begin{align}
\label{metrics}
\begin{split}
     & ds_{(+)}=-f(r)_{+}dt^{2}+\frac{dr^{2}}{f(r)_{+}}+r^{2}d\Omega^{2}_{2},\\
      & ds_{(-)}=-f(r)_{-}dt^{2}+\frac{dr^{2}}{f(r)_{-}}+r^{2}d\Omega^{2}_{2}.
\end{split}
\end{align}
Choosing $\Sigma$ to be an $r=r_{0}$ hypersurface, 
enforcing (J1) results in  $f(r_{0})_{(+)}=f(r_{0})_{(-)}$. The induced line element is $ds^{2}_{3}=-f(r_{0})dt^{2}+r_{0}^{2}d\Omega^{2}_{2}$ with normal $\sqrt{f(r)}^{ \ -1} dr$. This means that the discontinuity in extrinsic curvature across the surface will be  $[k_{ij}]=\frac{1}{2}\sqrt{f(r)}[\partial_{r}\gamma_{ij}]$. The metrics $g_{\mu \nu (+)}$ and  $g_{\mu \nu (-)}$ were chosen such that the matching of the angular components is trivial. Consequently, $[k_{\theta \theta}]=[k_{\phi \phi}]=0$ and the only non-vanishing component of the extrinsic curvature difference is $[k_{tt}]=\frac{1}{2}\sqrt{f(r)}[\partial_{r}\gamma_{tt}]$ due to $f(r)_{+}$ and $f(r)_{-}$ having different radial derivatives away from $\Sigma$ in general. The trace of the extrinsic curvature difference is then simply $[k]=\gamma^{tt}[k_{tt}]$, so that (J2) implies that the surface stress energy tensor $S_{ij}$ takes the form
\begin{equation}
\label{SS}
    S_{tt}=0 \quad  \& \quad S_{ab}=\frac{1}{8\pi}[k]\gamma_{ab},
\end{equation}
where $a,b$ are the angular coordinates on the 2-sphere (notice that with our notation, for our timelike surface, $i=t,a$, i.e. the intrinsic coordinates contain one timelike and two spacelike, angular components). 
Therefore $\Sigma$ is a zero energy density hypersurface that is characterised solely by a tangential pressure $p_{T}=\frac{1}{2}S$, where $S=S_{ij}\gamma^{ij}$ and the factor of $\frac{1}{2}$ accounts for the degeneracy of $p_{T}$ in the $\theta$ and $\phi$ directions. Hence 
\begin{equation}\label{pTk}
     p_T= \frac{[k]}{8\pi }. 
\end{equation}

In fact the absence of surface energy is obvious, seen from another angle. 
When $\Sigma$ is a thin-shell junction between two static and spherically symmetric geometries, the vanishing of $S_{tt}$ is required for the internal consistency of (J1) and (J2).  This can be eloquently argued from the Misner-Sharp energy \cite{MS}, cast in the form:
\begin{equation}
    E=\frac{r}{2}(1-g^{\mu \nu}\partial_{\mu}r\partial_{\nu}r),
\end{equation}
which is the most general definition of energy for spherically symmetric geometries \cite{Hayward1}. In a similar fashion to the other tensorial quantities, we may define the quantity $[E]\equiv E_{(+)}-E_{(+)}$ which represents the difference between the Misner-Sharp energies in $\mathcal{M}_{(+)}$ and $\mathcal{M}_{(-)}$. If $[E]\neq 0$, there is a contribution to the overall energy from the surface $\Sigma$. The explicit form of this difference for our construction is $[E]=\frac{r}{2}[g^{rr}]$. Utilizing the form of the metric from (\ref{metrics}) allows for the re-expression of this relation on $\Sigma$ as $[E]=\frac{r}{2}[\gamma_{tt}]$. In this form it is apparent that a non-zero energetic contribution from $\Sigma$ directly contradicts (J1).

The previous argument is completely general, regardless of what is put inside (or outside), as long as the junction condition J1 can be satisfied. Given the form of \eqref{metrics}, some extreme cases can be excluded, for example two Schwarzchild solutions with different masses. However, this still leaves us with a vast array of options, the most obvious one being a deSitter (dS) or anti-dS (AdS) interior glued to empty space.  For the former, we
can explicitly evaluate the solution to find a positive $p_T\sim \Lambda r_0/\sqrt{1-\Lambda r_0^2}$ capable of balancing the inwards tension due to a positive $\Lambda$. For an AdS interior glued to an empty exterior, the surface would be tense instead, $p_T<0$, for equivalent reasons, and similar sign results could be found for more general interiors and exteriors. This agrees with previous discussions for the gravastar~\cite{MottolaReview}. The sign of $p_T$ will have important consequences for our thermodynamical discussion.

\section{GENERAL THERMODYNAMICAL ARGUMENT}\label{thermodynamics}

This leaves us with a rather unwonted situation in thermal physics: a zero-energy system. Let us first see what general thermodynamical arguments (e.g.~\cite{MTW,brown,Ball}) can teach us about our peculiar matching surface, devoid of energy but with pressure and presumably other thermodynamic quantities. We start with the analogue of 
the first law of thermodynamics, initially without imposing vanishing energy. Assuming that there are no conserved particle numbers, this reads 
    $dU=TdS-p_{T}dA 
    $
where $U$, $S$ and $T$ are energy, entropy and temperature and $A=4\pi r^2$ is the area. 
By defining the associated intensive energy and entropy areal densities ($\rho = U/A$, $s=S/A$)\footnote{Notice that our $s$ is {\it not} a density per unit of particle, as in \cite{MTW,brown}, something that would be ill-defined, since we are assuming no conserved particle numbers.}
and substituting into the expression for $dE$ we get
\begin{equation}\label{Gibbsintensive}
    \bigg(\rho+p_{T}-Ts\bigg)\frac{dA}{A}=Tds-d\rho.
\end{equation}
We can generically consider the energy representation \cite{Ball} in which $\rho=\rho(s)$, so that $d\rho=\left. \frac{\partial \rho }{\partial s}\right|_A ds=Tds$. Then, the right hand side of \eqref{Gibbsintensive} vanishes, and 
we obtain the equivalent of the Euler relation for the surface $\rho +p_{T}=Ts$. Imposing now $\rho=0$ upon this general expression we finally get the equivalent of the Euler relation for a zero-energy surface:
\begin{equation}
\label{euler}
    p_{T}=Ts.
\end{equation}
As a corollary we get for the Helmoltz free energy areal density ${\cal F}=F/A=(U-TS)/A$:
\begin{equation}
\label{Fsigma}
    {\cal F}=-Ts=-p_T.
\end{equation}
This has to match the calculation obtained via the euclideanization of the path integral for the fluid. However, to this we should add the gravitational path integral. We will deal with both matter and gravity path integrals together in the next Section.

General thermodynamics, can take us further. Indeed $\rho=0$ can be seen as an equation of state, and this contains in itself other equations of state, up to integration constants. Using the Maxwell relations:
\begin{equation}
    \frac{\partial^2 F}{\partial A \partial T}=
  \frac{\partial^2 F}{\partial T\partial A} \implies
  \frac{\partial p_T}{\partial T}= \frac{\partial S}{\partial A}
\end{equation} 
in combination with the Euler relation \eqref{euler}, we get:
\begin{align}
  p_T&=\alpha T\label{eqstate}\\
      s&=\alpha\label{salpha}
\end{align}
where $\alpha>0$ is a constant with units of inverse area.
This defines a fundamental length, which could be build from $G$ (and so be the Planck length) or not. 
This equation of state fixes the temperature from the $p_T$ worked out from junction condition J2.

Given that $p_T$ can be positive or negative, and that $s$, related to the number of states, is positive, we learn that we can have negative temperatures in some circumstances but not in others. Specifically, a deSitter interior glued onto an empty exterior, in the spirit of a gravastar, implies a positive temperature on the surface, but an AdS interior would require a negative temperature. This interesting violation of the third law of thermodynamics (at least in one of its formulations) has parallels with nuclear spin systems in a magnetic field.



\section{The Euclidean Path Integral}\label{Sec:Z}
To calculate the thin shell partition function $\cal{Z}$$_{\Sigma}$, we start with the gravitational action proposed in~\cite{RC}:
\begin{equation}
    I_{g\Sigma}=-\frac{1}{8\pi}\int_{\Sigma} d^{3}x\sqrt{-\gamma} \ [k],
\end{equation}
which is simply the difference of the Gibbons-Hawking-York boundary terms associated with the actions on the regions  $\cal{M_{(+)}}$ and $\cal{M_{(-)}}$ on both sides of $\Sigma$. As shown in~\cite{RC}, the combined variation of this action and the matter action on $\Sigma$ yields the junction condition J2. The matter action can be written as\footnote{This reduces to the Nambu-Goto action for a domain wall with $p_T=-\mu$, but also to the action of a scalar field with generic kinetic and potential terms, $K$ and $V$. For perfect fluids there is more ambiguity in defining the on-shell value of the action, but the same result is a possible choice.}:
\begin{equation}
    I_{m\Sigma}=\int_{\Sigma} d^{3}x\sqrt{-\gamma} p_T.
\end{equation}
Using $I_\Sigma=I_{m\Sigma}+I_{g\Sigma}$, we define 
the path integral ${\cal Z}_\Sigma$ associated with the thin shell as $\mathcal{Z}_{\Sigma}=\int \exp[iI_\Sigma]$.
Upon eulideanization this provides the partition function $Z$.

\subsection{Definitions of temperature}
In general 
the partition function is expected to be a function of the inverse temperature $\beta$. In the presence of a horizon, one may infer $\beta$ from the fact that the space-time geometry near the horizon is a Rindler geometry. Wick rotating to Euclidean time $\tau$ via $\tau=it$ and enforcing the absence of conical singularities leads to the identification $\tau \sim \tau+\beta$ where $\beta$ is the inverse temperature of the horizon. In the absence of a horizon, however, there is considerable ambiguity concerning whether or not the temperature can be purely geometric, as we now discuss.

In the Gravastar construction \cite{gravastar1, gravastar2}, the aforementioned difficulties concerning temperature were resolved by considering a {\it thick}-shell junction between geometries being joined. To characterise the fluid in the thick shell, the Stefan-Boltzmann law,  $\rho \propto T^{d+1}$ and the equation of state for an isotropic perfect fluid, $p=\frac{1}{d}\rho$ in $d$ dimensions were invoked. Specifically, by setting $d=1$ the fluid is defined by the relations $\rho=p$ and $\rho \propto T^{2}$. However, an important consequence of this characterisation is that the shell now has an associated Misner-Sharp energy due to the presence of a non-vanishing energy density $\rho$. 

In keeping to the infinitely thin shell limit, we must therefore avail ourselves of another, more geometrical prescription of temperature for a preferred time-like surface. Let us sub-foliate $\Sigma$ into spheres and a timelike direction, recasting the induced metric on $\Sigma$ in the form $ds^{2}_{3}=-N^{2}dt^{2}+r_{0}^{2}d\Omega_{2}^{2}$, where $N=\sqrt{f(r_{0})}$ is the lapse function of the subfoliation into space-like surfaces $^{2}B$, with induced metric $ds^{2}_{2}=r_{0}^{2}d\Omega_{2}^{2}$ and a time-like normal $-Ndt$. Given this time-like normal, in~\cite{TE,BY,CR} one defines the inverse Tolman-Ehrenfest temperature from:
\begin{equation}
    \beta_{TE}=\int_{t'}^{t''} dt \  iN \bigg|_{\Sigma} .
\end{equation}
In \cite{BY}, the complexified lapse function $N_{C}\equiv iN$ is integrated over the real time parameter $t$, but, alternatively, one can choose to keep the lapse function real and integrate over thermal time $\tau$. Both options may be considered as Euclidean and in both instances the geometric interpretation of the integral is the same. The inverse temperature $\beta_{TE}$ represents the proper time that elapses between the space like surfaces $^{2}B$  as measured along curves in $\Sigma$ for both real and thermal time parameters. This is just one possible definition of temperature in this context, but it is worth noting that this definition is essentially equivalent to the original definition of the Tolman-Ehrenfest temperature defined from $T_{TE}\sqrt{g_{\mu \nu}\xi^{\mu}\xi^{\nu}}={\rm constant}$, where $\xi=\xi^{\mu}\partial_{\mu}$ is the time-like killing vector associated with the static geometry~\cite{TE,CR}.

\subsection{The Euclidean path integral}
We now show that this is precisely the definition of temperature required for making sense of the thermodynamical arguments in Section~\ref{thermodynamics}
from a path integral perspective, and to complete them with the gravitational component of the path integral. Upon Wick rotation, the path integral becomes the partition function $Z_{\Sigma}=\int \exp[-
I^E_{\Sigma }]$, where $I^E_{\Sigma}$ is the euclideanised action on $\Sigma$.
This contains both a gravity and a matter components, with:
\begin{align}
     I^E_{g\Sigma}&=\frac{1}{8\pi}\int_{\Sigma} d^{3}x\sqrt{-\gamma} \ [k],\nn\\
       I^E_{m\Sigma}&=-\int_{\Sigma} d^{3}x\sqrt{-\gamma} p_T.\label{Eucaction}
\end{align}
The solutions are $p_T=[k]/(8\pi)$ as in the Lorentzian case.
In general, one finds the equilibrium free energy
via the saddle point approximation, which amounts to replacing the solutions into the action (i.e. evaluating the ``on-shell'' value of the action). For $\Sigma$ we have $Z[\beta]\approx \exp[-I^{E}_\Sigma[\gamma_{ij}^{0},p_T^0]]$ where $\gamma^0_{ij}$ and $p_T^0$ satisfy the Einstein equations $p_T=[k]/(8\pi)$. It then follows for the free energy that $\beta F_\Sigma=I^E_\Sigma[\gamma_{ij}^{0},p_T^0]$.

The euclideanised volume element on $\Sigma$ is $\sqrt{\gamma}=N_{C}\sqrt{\sigma}$ where $\sqrt{\sigma}$ is the area element on a 2 sphere and, we recall, $N_C=iN$ is the complexified lapse function for the subfoliation of $\Sigma$ into spheres and a timelike direction. Hence the 3D volume integral to which both $I^E_{g\Sigma}$ and $I^E_{m\Sigma}$  are proportional is $V_3=\beta_{TE}A$, where $A=4\pi r_0^2$, giving  confirmation that the Tolman-Ehrenfest temperature is the correct one to consider here. Bearing this in mind, and focusing now on the matter component, we obtain 
\begin{equation}
    -\ln Z_{m\Sigma}=\beta_{TE} F_{m\Sigma}=I_{m\Sigma}^E=-V_3 p_T=-\beta_{TE}Ap_T.\nn
\end{equation}
Hence the matter free energy, with this definition of temperature, comes out as:
\begin{equation}
    F_{m\Sigma}=\frac{I_{m\Sigma}^E}{\beta_{TE}}=-p_TA
\end{equation}
implying the thermodynamical prediction \eqref{Fsigma}.
This is negative for $p_T>0$, corresponding to a positive temperature, for example for a dS interior matched onto an empty exterior (and vice versa for an AdS interior).

The other thermodynamical predictions in Section~\ref{thermodynamics} follow from this partition function, once one adapts to the fact that $U=\langle E\rangle=0$. The differences are listed in table~\ref{table}.
\begin{table}[h!]
\centering 
\begin{tabular}{|c|c|c|} \hline
\makecell{Quantity of \\ interest} & \makecell{Standard \\ canonical ensemble} & \makecell{ Our surface $\Sigma$}\\ \hline 
 \makecell{Partition function  \\ $Z$} &
 $ e^{-\beta F}$ & $ e^{S}$\\ \hline 
 \makecell{Entropy \\ $S$} & 
 ${(1-\beta \partial_{\beta})\ln Z}$ & 
 {$\ln {Z}_{\Sigma}$}\\ \hline 
 \makecell{Energy \\ $U=\langle E \rangle$} &
 {$-\partial_{\beta}\ln Z$} & 0 \\ \hline 
\makecell{Pressure \\ $p$ or $p_T$} & 
{$ T\partial_{V}(\ln Z)$} & 
{$ T_{TE} \partial_{A}(\ln Z_{\Sigma}$)}\\ \hline 
\end{tabular}
\caption{Partition function and thermodynamic quantities in the standard canonical ensemble and 
for our zero energy surface.}\label{table}
\end{table}
The crucial difference is that the in the saddle point approximation the matter partition function is approximated by the unusual:
\begin{equation}
    Z\approx e^S 
\end{equation}
and does not actually depend on the temperature. This must be true, so that here is no contradiction between the usual formula for $U$ and the fact that $U$ is zero on $\Sigma$. The formula for the entropy and pressure are also the usual ones, reduced to our specific case, and they simply imply the Euler relation \eqref{euler} (as long as the Tolman-Ehrenfest temperature is used). Finally it is easy to derive the equation of state \eqref{eqstate} from the fact that $\ln Z=S=p_T A \beta_{TE}$ and that this cannot depend on $\beta_{TE}$, so that $U=0$. Hence $p_T=\alpha T_{TE}$. The expression $S=\alpha A$ then also follows, and confirms \eqref{salpha}.

We have thus confirmed the matter thermodynamics obtained in Section~\ref{thermodynamics} from a Euclidean path integral perspective. Adding gravity, however, reveals a vanishing total free energy for $\Sigma$, at least to leading order in the saddle point approximation. This is because the total Euclidean action vanishes on shell, as can be readily checked by inserting $p_T=[k]/(8\pi)$ in~\eqref{Eucaction}. 
Not only does $\Sigma$ have no energy $S_{tt}$; it does not have any free energy once the gravity component is taken into account. This is to be contrasted with other similar calculations (for example for a standard Nambu-Goto membrane~\cite{Caldwell}).

Nevertheless, the matter content in $\Sigma$ has a temperature, a free energy and entropy, fixing the range of integration in imaginary time, and thus the temperature everywhere in the space, once the Tolman-Ehrenfest effect is taken into account. 

\section{Conclusions}

We kept arguments general, but in one of its applications the construction in this paper can be seen as an extreme gravastar, with its ``atmosphere'' compressed into a membrane (which does not need to be near the horizon in our general case). In keeping with the gravastar (whose atmosphere contains negligible energy but most of the entropy) in our construction the timelike surface contains precisely no energy, and potentially all of the {\it thermodynamical} entropy of the space.

The thermodynamic entropy is the area $A$ of the surface in units of $1/\alpha$, which does not depend on the temperature and could be the Planck area.  So there is some parallel with the Bekenstein result, even though the surface is not a horizon. The temperature on $\Sigma$ must be the Tolman-Ehrenfest temperature, and is proportional to the transverse pressure on the surface, itself proportional to the jump in the trace of the extrinsic curvature across the surface. Therefore, the rough parallel with black holes is not confined to the entropy. 
Obviously here there is no Killing horizon to support the usual definition of surface gravity (nor does our definition of extrinsic curvature apply to null-surfaces), but the trace of the extrinsic curvature is a good definition of surface gravity for our time-like surface (it reduces to it in the Newtonian limit, $K\sim |\partial _r\Phi_N|$, with $f\approx 1+2\Phi_N$ and $\Phi_N\ll 1$). Hence the transverse pressure and the temperature are proportional to the jump in surface gravity across the surface $\Sigma$, reinforcing a rough analogy with the black hole case. 

And yet, in spite of these analogies, this is fundamentally not the usual result: the entropy and free energy are purely from the matter components, and the surface is not a null surface (so the concept of surface gravity is intrinsically different).
The results in this paper are intriguing and merit further developments.

\section{Acknowledgments}
We thank Niayesh Afshordi, Andrew Tolley and Toby Wiseman for serious discussions and arguments. RI was supported by a Bell-Burnell Fellowship and JM agtly supported by STFC Consolidated Grant ST/T000791/1.

\end{document}